\documentclass{aa}
\usepackage{graphicx}
\nonstopmode
\newcommand{\et}{{\it et al.}}

\begin{document}

   \title{Rapidly Evolving Circularly Polarized Emission during the 1994
   Outburst of GRO~J1655-40}

\authorrunning{Macquart, Wu, Sault \& Hannikainen}
\titlerunning{Circular polarization in 1655$-$40}

   \author{J.-P. Macquart\inst{1}, K. Wu\inst{2},
   R.J. Sault\inst{3} \and D.C. Hannikainen\inst{4,5}   }

   \offprints{J.-P. Macquart}

   \institute{Kapteyn Astronomical Institute, University of Groningen,
                Postbus 800, 9700 AV Groningen, The Netherlands\\
                \email{jpm@astro.rug.nl}
         \and
             Mullard Space Science Laboratory, University College London,
             Holmbury St Mary, Surrey RH5 6NT, United Kingdom, and
             School of Physics A28, University of Sydney, NSW 2006,
             Australia \\
             \email{kw@mssl.ucl.ac.uk}
        \and
             Australia Telescope National Facility, Narrabri, NSW 2390, Australia \\
                \email{rsault@atnf.csiro.au}
        \and
             Department of Physics and Astronomy, Southampton University,
             Southampton, SO17 1BJ, United Kingdom 
	\and
	    Observatory, PO Box 14, Fin-00014 University of Helsinki, Finland
		\email{diana@astro.helsinki.fi}
             }


\abstract{
We report the detection of circular polarization
  during the 1994 outburst of the Galactic microquasar GRO~J1655$-$40.
The circular polarization is clearly detected at 1.4 and 2.4~GHz, but not at 4.8 and 8.4~GHz, where its magnitude never exceeds 5~mJy.
Both the sign and magnitude of the circular polarization evolve
  during the outburst.
The time dependence and magnitude of the polarized emission
  can be qualitatively explained by a model
  based on synchrotron emission from the outbursts, but is most consistent with
  circular polarization arising from propagation effects through the
  relativistic plasma surrounding the object.
}

   \maketitle
%

\section{Introduction}

GRO~J1655$-$40 (Nov Sco 1994) is a transient X-ray binary at an estimated distance of 3.2\,kpc (Hjellming \& Rupen 1995) 
   containing a low mass F3IV--F6IV companion star 
   and a black-hole candidate which has a mass about 7~M$_\odot$  
   (Orosz \& Bailyn 1997; Soria \et\ 1998; Shahbaz \et\ 1999).    
It was first detected in the hard X-ray band   
   by the Burst and Transient Source Experiment (BATSE)  
   on board the {\it Compton Gamma-Ray Observatory}   
   on 1994 July 27, at the onset of an X-ray outburst (Zhang \et\ 1994).  
Its X-ray flux in the 20--100~keV band quickly reached 1.1 Crab 
   on August 1.   
The system remained in an outburst state until about August 15   
   and then was briefly quiescent,  
   but flared again on September 6 (Harmon \et\ 1995).
GRO~J1655$-$40 underwent several further outbursts between 1994 and 1997
   before retreating into a quiescent state in late 1997.

The radio counterpart of GRO~J1655$-$40 was first identified  
   by the Molongo Observatory Synthesis Telescope (MOST)  
   at 843~MHz (Campbell-Wilson \& Hunstead 1994a)  
   shortly after the BATSE discovery.  
The radio flux density showed a steep initial rise,   
   and it reached 4.2 and 5.5~Jy on 1994 August 14 and 15 respectively  
   (Campbell-Wilson \& Hunstead 1994b).   
Imaging using the Very Large Array (VLA), the Very Long Baseline Array (VLBA),  
   and the Southern Hemisphere VLBI Experiment (SHEVE)  
   showed repeated episodes of relativistic ejections from the system  
   (Hjellming \& Rupen 1995; Tingay \et\ 1995),  
   with the ejecta in superluminal motion.  
After correction for the inclination to the line of sight,  
   the ejection velocity was inferred to be $v\sim 0.92c$
   in the rest frame of the system (Hjellming \& Rupen 1995).

During the 1994 ejection events, GRO~J1655$-$40 was monitored  by the MOST,
   the Australian Telescope Compact Array (ATCA)  
   and the Hartebeesthoek Radio Astronomy Observatory (HartRAO)  
   (see Hannikainen \et\ 2000 for details).  
The radio emission showed strong (up to about 10\%) linear polarization.  
The variations in the multi-frequency linear polarization light curves  
   reveal several ejection events and strong opacity effects.   
The simultaneous peaking of multi-frequency radio light curves at about August 19 
   contradicts the prediction of the conventional synchrotron bubble models  
   (e.g.\ Hjellming \& Johnston 1988).  
This behaviour is, however, consistent  with that of the generalized-shock model 
  for AGN jets  (Marscher \& Gear 1985),  
  for the shocks in the jets of GRO~J1655$-$40 in the initial growth stage  
  (Wu, Stevens \& Hannikainen 2002; Stevens \et\ 2002).

Here we report the detection of circular polarization in the radio emission from GRO~J1655$-$40.
Circular polarization has been detected in two other Galactic jet sources: SS433 (Fender \et\ 2000) and GRS~1915$+$105 (Fender \et\ 2002).  For SS433 the observations contained data from two epochs only; for GRS~1915$+$105 the observations were conducted only after (as opposed to during) the onset of the ejection episodes and their associated radio emission.

Circular polarization is also observed in a broad variety of AGN, 
  from relatively weak sources like Sgr~A* and M81*  
  (Bower, Falcke \& Backer 1999; Sault \& Macquart 1999; Brunthaler \et\ 2001) 
  to powerful quasars such as PKS~1519-273, 3C~273 and 3C~279  
  (Wardle \et\ 1998; Homan \& Wardle 1999; Macquart \et\ 2000).  
The origin of the circular polarization is uncertain (Wardle \& Homan 2001; 
Macquart 2002). 

The two most likely causes of the circular polarization in Galactic jet sources are the small amount intrinsic to synchrotron radiation, and that which results from the propagation of linearly polarized radiation through a relativistic plasma (Legg \& Westfold 1968; Pacholczyk 1973; Cheng, Pacholczyk \& Cook 1985; Fender \et\ 2002). 
Identification of its cause and relationship to other observable jet parameters can constrain, for instance, the properties of the magnetic field in the jet and the low energy end of the relativistic electron distribution.  This in turn may indicate whether the composition of the relativistic jets is primarily leptonic or baryonic (Wardle \et\ 1998; but see also Ruszkowski \& Begelman 2002).

In determining the origin of the circular polarization, it is pertinent to compare the properties of circular polarization observed in AGN with that in Galactic jet sources.  The magnitude of the circular polarization is similar in both classes of source.  Despite this, it has not been possible to identify the origin of the circular polarization in either SS~433 or GRS~1915$+$105.

Here we report variable circularly polarized emission during the 1994 August--September outbursts of GRO~J1655$-$40.  The spectral and temporal coverage of the circular polarization over the duration of the flaring in this source presents an exceptional opportunity to quantify the properties of the circular polarization in a Galactic X-ray binary.  In particular, this is the first time that the circular polarization has been measured during the commencement of the radio flare, during which the source appears to exhibit dramatic polarization variability.

We discuss the observations in \S2 and present a model-independent interpretation of the data in \S3.  These results are discussed in the context of a model for the outbursts in GRO~J1655$-$40 in \S4, and the characteristics of the circular polarization in this source are related to those of other 
X-ray binaries and AGN.   The conclusions are presented in \S5.


\section{Observations and Results}
\subsection{Data reduction}

The data presented in this paper were obtained with ATCA over the interval 1994 August 15 to 1994 September 3 (TJD = 9579 to 9597).  The ATCA is an Earth-rotation aperture synthesis array, comprising six 22~m antennas which can be moved along an east-west track to give baselines up to 6~km (Frater, Brooks \& Whiteoak 1992).  The observations of GRO~J1655$-$40 were made at centre frequencies of 1.380, 2.378, 4.800, 5.904, 8.640 and 9.200\,GHz with 128-MHz bandwidth in two orthogonal polarizations.   Each observation was typically 10 minutes in duration, except on August 15, where the observations spanned 3 hours at 1.380 and 2.378\,GHz and 5 hours at 4.800 and 8.640\,GHz.  The antenna gain and phase calibration were derived from regular observations of the point source calibrator PKS~1740$-$517.  The flux density scale was tied to the ATCA primary flux calibrator PKS~1934$-$638 (Reynolds 1994).  The array was in the 6.0A configuration throughout these observations, giving interferometer spacings from 337 to 5939\,m.  
The data from the bands centered at 5.904 and 9.200\,GHz have been excluded from the present discussion, since the stability of the instrumental response to circular polarization is not well characterized at these frequencies.

Data reduction was performed in the {\small MIRIAD} package (Sault, Teuben \& Wright 1995).   
The instrumental leakages and antenna  gains were determined simultaneously using either PKS~1934$-$638 or PKS~1740$-$517.  At all wavelengths a single round of phase self-calibration was performed on the GRO~J1655$-$40 data (using a point source model) to eliminate residual phase instability.

The reduction of this dataset for variations in the total and linearly polarized flux density is discussed further in Hannikainen \et\ (2000).  However, some specific comments relating to the circular polarization are in order.

\begin{figure*}
\centering
\includegraphics[width=10cm]{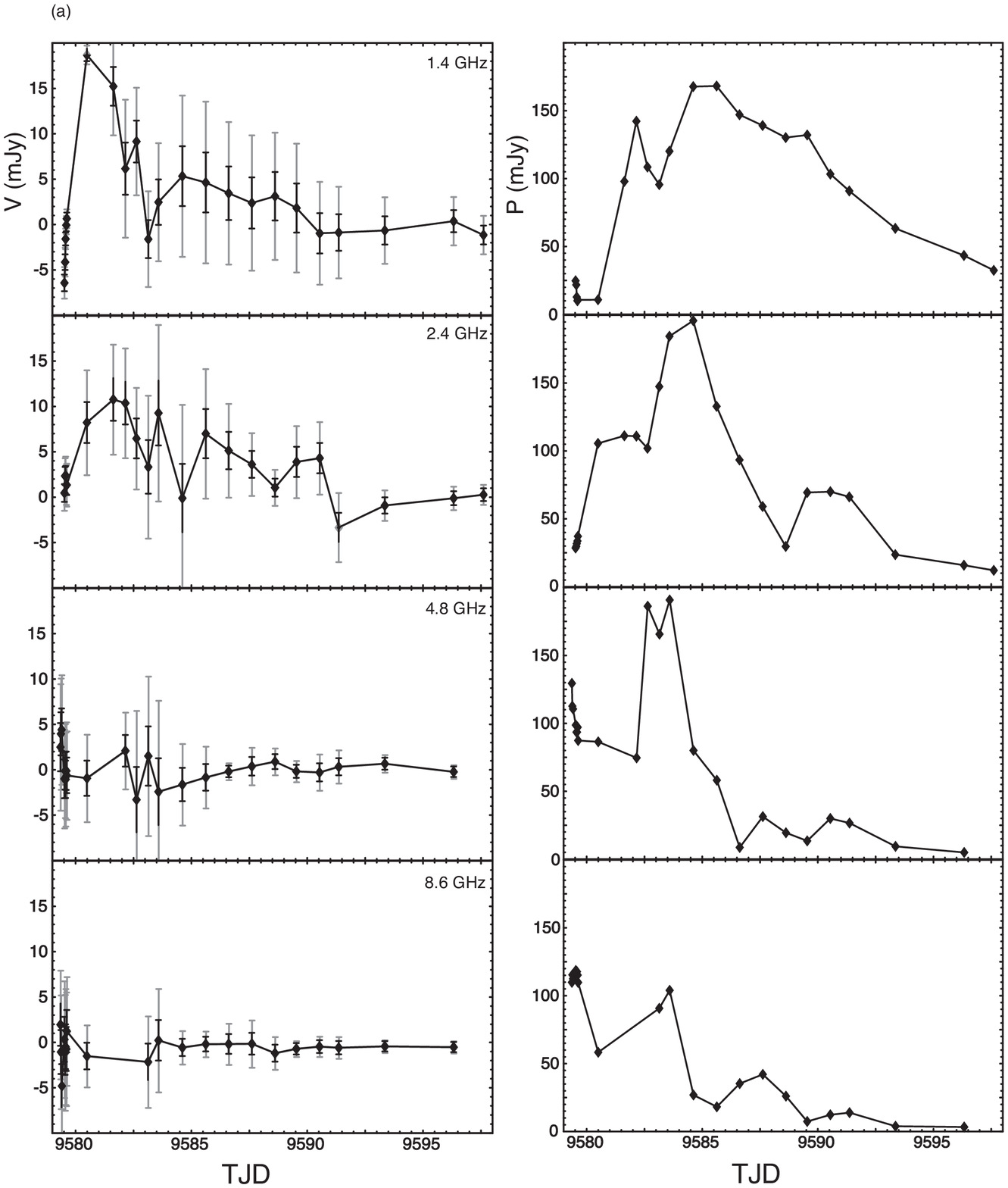}
\includegraphics[width=10cm]{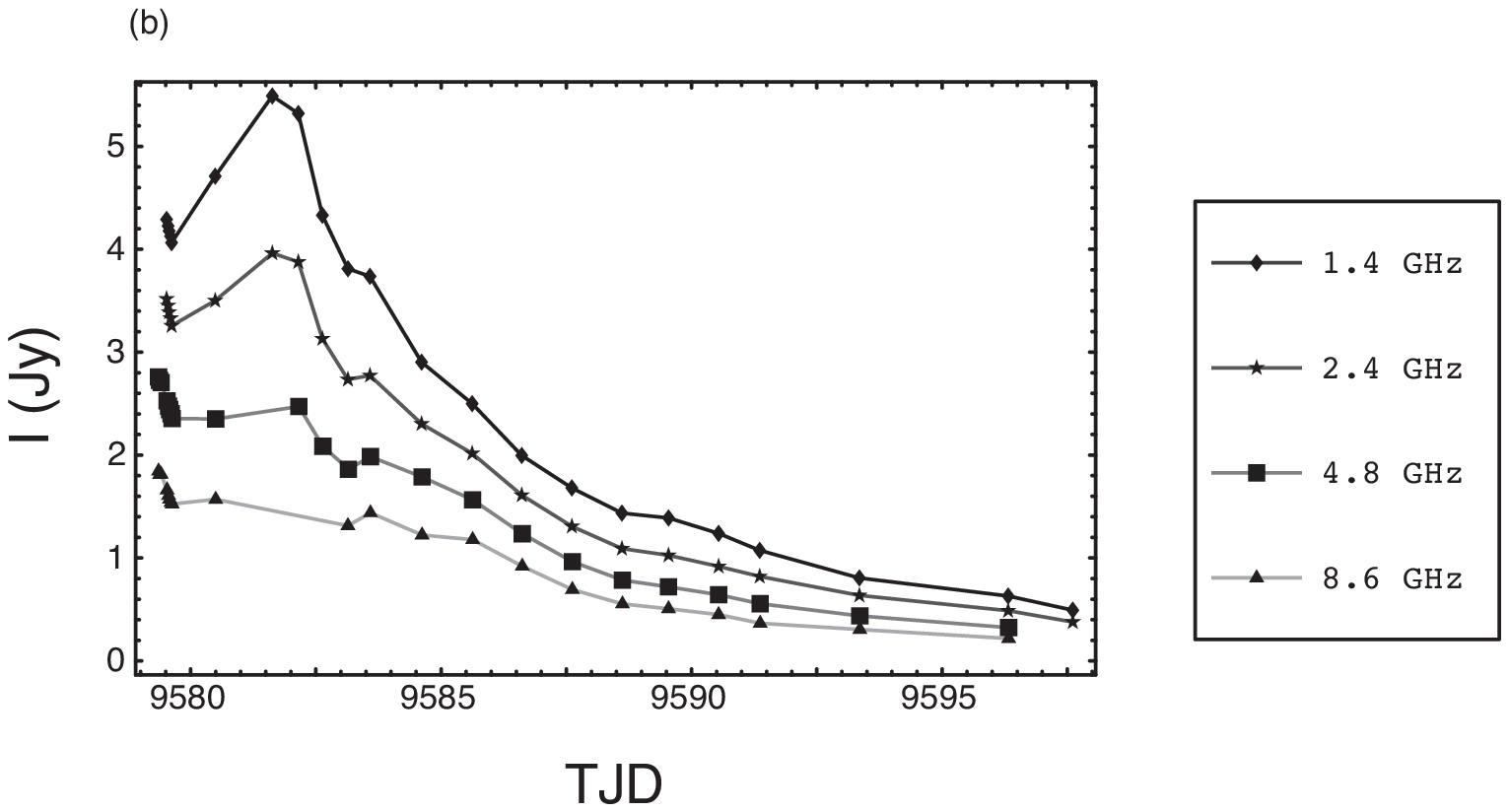}
\caption{
(a)  Evolution of the circular (left) and linear (right) polarization in GRO~J1655$-$40 at 1.4, 2.4, 4.8 \& 8.6\,GHz (TJD=JD$-$2,440,000.5).   The black error bars for the circular polarization values represent the combination of 1$\sigma$ thermal noise errors with the one third of the maximum possible error due to antenna leakage uncertainties; as such the black error bars should be interpreted $1\sigma$ error bars in both random and systematic error.
The gray error bars represent the 1-$\sigma$ thermal noise errors combined with the {\it maximum} possible uncertainty due antenna leakage, which is $0.05 \times {\cal P}$ (see text).  The circular polarization variations at TJD~9579 at 1.4\,GHz are reproduced in greater detail in Figure 2. The linear polarization plots are adopted from Hannikainen \et\ (2000), and the errors in these curves are $\sim 1$~mJy.  (b) Lightcurves of the variations in Stokes $I$ at the same four frequencies.} 
\label{VStokesFig}
\end{figure*}  

Normal polarization calibration of ATCA data assumes the `weakly polarized' case (Sault, Killeen \& Kesteven 1991), where the terms in `$leakage \times (Q,U,V)$' are neglected.  The `weakly polarized' case does not calibrate leakage between linearly and circularly polarized emission. 
The {\small MIRIAD} calibration routines allow solution using the `strongly polarized' case, which also solves for second order terms.  To determine such a calibration solution requires a calibrator source which is at least a few per cent linearly polarized, so that there is sufficient signal in the second order terms.

In the present case, the lack of observation of an appropriate linearly polarized source forced the use of the weakly polarized solutions to determine the instrumental leakage. 
This presents a limitation to the polarimetric accuracy of the circular polarimetry; the ATCA can measure circular polarizations to a level $V/I < 0.01$\% (Rayner, Norris \& Sault 2000) provided the leakages are determined using the strongly polarized solutions.  However, extensive comparison against the strongly-polarized solutions indicates that use of the weakly polarized equations results in an uncertainty in Stokes $V$ {\it at worst} 5\% of the total linearly polarized flux density (Rayner 2000).  The level of the circular polarization detected in GRO~J1655$-$40 is sufficiently high relative to the linearly polarized flux density (hereafter denoted ${\cal P}$) that we are confident of the reality of the circular polarization in this source.  We further note that polarimetric leakages are quite stable over time scales of several days on the ATCA.


We have adopted a limit of $0.05\,{\cal P}$ due to polarization leakage as the total absolute error in quoting errors for the circular polarization.  
However, several arguments suggest this limit is conservative.

The linear polarization does not follow the same trends as the circular polarization. 
In order to explain a substantial fraction of the variations in Stokes $V$ as an effect of instrumental leakage, this  would require large {\it variations} in the degree of polarization leakage on the time scale of the observed variability, i.e. from day to day.  This is at variance with the fact that leakages are known to be stable on ATCA over periods exceeding several days.

Some specific limits are provided by detailed examination of the data.
\begin{itemize}

\item 1.4\,GHz: The circular polarization measured between the two days TJD~9581 and 9582 are of comparable magnitude, even though the linear polarization increased sharply between these  two epochs. The linearly polarized flux density is $<10\,$mJy on TJD 9581,
while the circularly polarized flux density is $18\,$mJy with a $0.5\,$mJy error due to thermal noise.  However, when the linear polarization increases to 150\,mJy on TJD~9582, the circular polarization is measured at 15~mJy.  Thus, {\it if} the variation in $V$ over this time interval is entirely due to instrumental leakage, the small change in $V$ compared with the large change in linear polarization implies that instrumental leakage contaminates the circular polarization at a level less than $0.02\,{\cal P}$.  

Moreover, we note that the general increase in linear polarization in the interval TJD~9581--9586 is not mirrored by an increase in the circular polarization.

\item At 2.4 GHz the linear polarization reaches 194~mJy on TJD~9584 but $V$ does not greatly exceed $10$\,mJy over the entire period of observation.  The leakage is thus smaller than 5.5\% of the linearly polarized flux density.

\item At 4.8 GHz the linear polarization reaches 191\,mJy but at no stage during or after the flaring does $V$ exceed 5\,mJy.  This suggests instrumental calibration results in a leakage of less than 2.3\% of the linear polarization into circular polarization.

\end{itemize}

\subsection{Results}  

Figure 1 shows the circular polarization and linear polarization from 1.4 to 8.6~GHz during the flaring.

The circular polarization lightcurves at 1.4 and 2.4~GHz are characterised by rapid changes over the first three days (TJD 9579$-$9582).  At 1.4~GHz the circular polarization changes from $-6.4 \pm 1.7$~mJy to $+0.6 \pm 1.0$~mJy over a period of 150 minutes at the start of the outburst, on TJD~9579 (see Fig. 2).  The sign change is confirmed on the following day, when the circular polarization is measured at $18.0 \pm 1.0$~mJy.  Less pronounced variability is observed at 2.4~GHz over the corresponding interval, and there is no evidence for a sign change during the initial outburst at this frequency.  At both frequencies the circular polarization then rapidly decreases from this maximum value, but shows a relatively sustained level of $\approx 4$~mJy from TJD 9584--9588.

Several marginally significant dips are apparent in the circular polarization lightcurves.  There is a marginally significant dip at 2.4~GHz on TJD~9591.36, where the circular polarization is negative before it returns to zero.  The detections of other `dips' in the circular polarization lightcurves, at 1.4~GHz at TJD~9582.15 and 9583.14 and 2.4~GHz at TDJ 9583.14 and 9584.61, are not significant if the telescope leakage is as high as 5\%, and are  marginally significant if the leakage is closer to 2\% (see \S2.1).

There is no clear detection of circular polarization at 4.8 or 8.6~GHz.  
The circular polarization never exceeds 5~mJy at either frequency.  
It is only possible to set restrictive upper limits on the circular polarization at late epochs 
when the degree of linear polarization is low.  After TJD~9589 the circular polarization is consistent with zero, with measured values between $0.6\pm 1.0$~mJy and $-0.3 \pm 2.0$~mJy at 4.8~GHz and between $-0.7\pm 1.5$ and $-0.5 \pm 1.5$~mJy at 8.6~GHz.

Figure 1 shows that several large changes in the 1.4 and 2.4~GHz circular polarization lightcurves bear a close resemblance to the flares identified in the linear polarization by Hannikainen \et\ (2000), as marked.  For instance, the rise and decay of the first circularly polarized outburst shows a corresponding burst in the linear polarization over the same epoch, from TJD~9580--9583.

While some features in the circular polarization lightcurves at 1.4 and 2.4~GHz appear to be connected to some in the linear polarization, the shapes of the lightcurves are qualitatively different.  This suggests that the fine structure of the lightcurves observed in $V$ is real, and not due to instrumental leakage from linear to circular polarization.

\section{Analysis} 

Only a small fraction of the source emission appears to be circularly polarized.   
The circular polarization lightcurve -- like the linear polarization lightcurve -- shows features which are not reflected in the total intensity.  Many of the polarized `outbursts' therefore contribute little to the total source emission. 

One can estimate the degree of polarization at 1.4~GHz on TJD~9579 if the rapid changes in $V$ are connected with those in $I$.  By comparing the rapid changes in the polarized and total intensity over this interval, the circular and linear polarizations of the variable component are estimated to be $-3.3 \pm 0.4$\% and $9.8 \pm 1.5$\% respectively at this frequency (see Figure 3).  We cannot estimate the degree of polarization at higher frequencies using this method because the changes in $V$ are not clearly detected.  Fender \et\ (2000) speculated that, although $\bar V/\bar I \sim 0.3$\% is measured in SS~443, the actual circular polarization of the polarized component was likely be an order of magnitude higher.  Our measurement substantiates this speculation for GRO~J1655$-$40.

It is impossible to explain the behaviour of the circular polarization in terms of the bursting and decay of a single right-handed circularly polarized component in the source.  We suggest that the evolution of the circular polarization is associated with the evolution of the multiple bursts emanating from the source, each of which is circularly polarized.  There is both transient right-handed {\em and} left-handed circularly polarized emission at various times.  
The negative circular polarization clearly detected at early epochs cannot reflect the quiescent level of the circular polarization since its level at later epochs, after the bursts, tends towards zero.  
Furthermore, the dip in the circular polarization at 2.4\,GHz (TJD 9591.36) also appears to be negative, suggesting that some component of the source `bursts'  in negative circular polarization.

We argue that the rapid decline of the (positive) circular polarization at around TJD~9581--9583 
is due to increased generation of negative circular polarization from another sub-component in the source, rather than decay of the component responsible for the positively polarized emission.  
The fact that the initial flare decays extremely rapidly at 1.4 and 2.4~GHz supports this.  Although the data is clearly marginal, the circular polarization also appears to rebound to higher (positive) values and decay more slowly after TJD~9584.5.


\begin{figure}
\centering
\includegraphics[width=6cm]{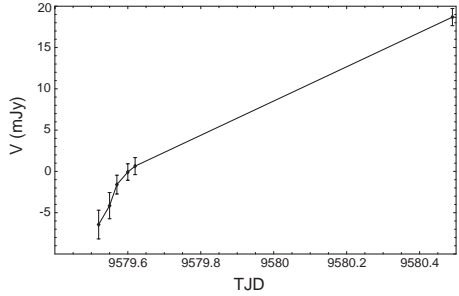}
\caption{
A blow-up of the lightcurve of Stokes $V$ at 1.4\,GHz during the onset of the radio outburst.  The errors in the values plotted are dominated by thermal noise, not antenna leakage uncertainties.} 
\label{Fig2}
\end{figure}

\begin{figure}
\centering
\includegraphics[width=6cm]{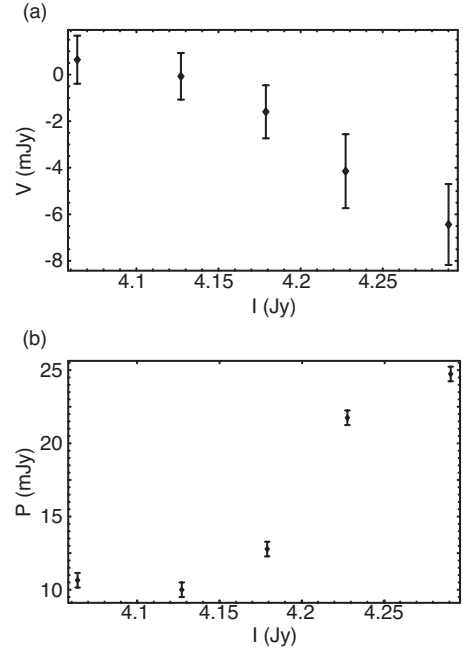}
\caption{(a) The relationship between $V$ and $I$ on TJD 9579 at 1.4\,GHz.
   If changes in the two quantities with time are related, 
       this implies the presence of a polarized {\it component} 
       of the source with $-3.3\pm0.4$\% circular polarization at 1.4\,GHz.
(b) The corresponding relationship between ${\cal P}$ and $I$ at 1.4\,GHz.  A fit to the earliest four  epochs (the rightmost four points) suggests a linear polarization of $9.8 \pm 1.5$\% in the varying component.  The point with $I=4.064\,$Jy corresponds to the last epoch on the first day of observations, at TJD 9579.62.}
\label{Fig3}
\end{figure}

\section{Discussion}

\subsection{Origin of Circular Polarization}

The circular polarization is most plausibly either intrinsic to synchrotron emission or arises as a result of polarization conversion through relativistic plasma associated with the source (see Pacholczyk 1977).  Other causes of circular polarization are less likely.  The angular diameter of the GRO~J1655$-$40 is too large for scintillation-based circular polarization (Macquart \& Melrose 2000) to be relevant here.  Emission from coherent emission processes is also unlikely, since the total flux density and linearly polarized emission are already well explained by synchrotron 
models (Hannikainen \et\ 2000; Wu \et\ 2002).

The circular polarization is negative at the onset of the radio flare on TJD~9579, but changes sign and peaks on TDJ~9580.  The spectrum is inverted at the low frequencies when the sign change of the circular polarization occurs at TJD~5979 and subsequently evolves to a pure power law after $\sim 1$~day (Hannikainen \et\ 2000).

In principle, both synchrotron radiation and polarization conversion are capable of explaining both the magnitude and sign changes observed in the circular polarization.  Both predict a change in the sign as the emission evolves from optically thick to thin (Melrose 1971; Pacholczyk 1977).

The degree of circular polarization expected from synchrotron radiation from particles with Lorentz factor $\gamma$ in a homogeneous magnetic field is $m_C \sim 1/\gamma$.   
If the synchrotron optical depth, $\tau$, exceeds unity at early epochs, as suggested by the data, the $-3.3$\% circular polarization deduced from the $V$ and $I$ variations on TJD~9579 could be due to the synchrotron emission from electrons with $\gamma \la 30$.  
The low degree of linear polarization, $m_P \approx 9.8 \pm 1.5$\%, deduced from the corresponding $P$ and $I$ variations is consistent with the value expected from optically thick synchrotron emission. For particles with energy distribution $N(\gamma) d\gamma \propto \gamma^{-s} d\gamma$, one expects $m_P = 3/(6 s+13)$ (e.g.\ Pacholczyk 1977).  
The linear polarization then implies $s \approx 2.9 \pm 0.9$.  This value is consistent with the spectral index of $-0.7$ estimated in Wu \et\ (2002) and Stevens \et\ (2002).  Lower values of $\gamma$ and $s$ are required if the magnetic field in the emission region is inhomogeneous and causes depolarization.

On the other hand, the emission may originate in an optically thin sub-component of the source.  One then needs to account for the discrepancy between the observed degree of linear polarization and the canonical value ${m_P}_{\tau < 1} =(s+1)/(s+7/3) \sim 70$\% expected for optically thin synchrotron radiation. 

The most likely explanation is that the observed emission is composed of contributions from several sub-regions, each with different intrinsic polarizations, caused by variations in the orientation of the magnetic field.  In this case the circular and linear polarizations would both be depolarized.  Thus the intrinsic circular polarization is higher than the measured value.  The intrinsic circular polarization would be of order $\sim 3.3 \times (70\% / 10\%)=23$\% if the depolarization affects both linear and circular polarization to a comparable degree.  This would require the circular polarization to originate from electrons with $\gamma \la 4$.

An external Faraday screen can also account for the (linear) depolarization, but this is unlikely since the measured  Faraday rotation in the source is only moderate: $\sim 20-100$~rad/m$^2$ (Hannikainen \et\ 2000). 




The circular polarization could result from the propagation of linearly polarized radiation through a relativistic plasma.  This circular polarization arises because the wave modes in a relativistic plasma are elliptical, and propagation through the plasma converts linear polarization to circular polarization, similar to the way in which Faraday rotation converts radiation between Stokes $Q$ and $U$ in a cold plasma.

This mechanism is easily capable of yielding circular polarizations of a few percent (Ruszkowski \& Begelman 2002).  Circular polarization produced via this effect is strongest for optical depths near unity.  As such, our data qualitatively favours this model, because the peak of the circular polarization at TJD~9580 coincides with the date at which $\tau \sim 1$ at 1.4\,GHz.  
The degree of circular polarization expected from synchrotron radiation for such optical depths does not exhibit a peak, but instead passes through zero (e.g.\ Pacholczyk 1977). 

Moreover, explanations based on synchrotron emission are not viable if the jet is composed mainly of e$^{\pm}$ pairs (see e.g., Kaiser \& Hannikainen 2002).  In this case little intrinsic circularly polarized emission is expected, as the circular polarization from each electron is equal in magnitude but of opposite sign to each positron.


\subsection{Circular polarization variations}

It is possible to explain the temporal variability of the circular polarization as a result of the emission from a succession of evolving ejections.

The circular polarization variability is qualitatively reproduced by a simple model in which the emission from the multiple ejections each evolves from optically thick to thin, and in which the circular polarization changes sign as it does so.  The circular polarization observed is the sum of the contributions from all previous ejections.

Hannikainen \et\ (2000) identify three separate ejection events in the period TJD 9579--9598 on the basis of the linearly polarized emission detected by ATCA at TJDs $\sim 9582,\ 9587$ and $9591$.

As the first burst commences it is initially optically thick and left-hand (negatively) circularly polarized.  The circular polarization changes sign as the emission becomes optically thin, as observed in the data.  However, at TJD~9582, a second burst occurs.  If it, too, is initially negative circularly polarized, this explains the rapid decrease in the apparent level of circular polarization at this epoch.  However, as this burst also becomes optically thin after $\sim 1$~day, the nett circular polarization returns to a high value, since the contributions from the two bursts are now both positive.

It is difficult to verify the model with data at later times because the high linear polarization causes large measurement uncertainties in the circular polarization due to instrumental polarization leakage.  However, the negative circular polarization at TJD~9591 at~2.4 GHz {\it may be} due to the emission from the third outburst.

This model has an obvious prediction. High resolution (VLBI) polarimetry capable of resolving individual ejecta at early epochs would detect circular polarization of order a few percent, and which reverses sign as the emission becomes optically thin to synchrotron radiation.

\subsection{Comparison with GRS~1915$+$105}

Fender \et\ (2002) report circular polarization associated with multiple outbursts in the Galactic superluminal source GRS~1915$+$105.  The circular polarization evolves on time scales of $\sim 3$ hours and is strongest at the lowest measured frequency, 1.4~GHz, as observed in GRO~J1655$-$40.

There is evidence for multiple components at varying optical depths; total flux density and spectral index lightcurves indicate there are at least four ejection events in the 2001 January outburst. 
The circular polarization is also reported to be coming from the parts of the system with `appreciable optical depth'.  The variability in the circular polarization could well be driven by opacity evolution in these components.  This is consistent with what is observed in GRO~J1655$-$40, where the circular polarization is also strongest at $\tau \sim 1$.

\subsection{Comparison with compact flat-spectrum AGN}

Circular polarization is detected in a number of AGN (e.g.\ Komesaroff \et\ 1984, Raynder, Norris \& Sault 2000).  In particular, it is found in the parsec-scale jets of blazars such as 3C~273, 3C~279 and 3C~84 (Homan \& Wardle 1999; Homan, Attridge \& Wardle 2001).  The circular polarization from these sources is highly variable.  We also note that scintillation variability measurements imply degrees of circular polarization about 4\% in some AGN (Macquart \et\ 2000; Macquart 2002), comparable to what we observe in GRO~J1655$-$40 (see Figure 3).

There is evidence that the degree of circular polarization is stronger in flat-spectrum sources (Rayner, Norris \& Sault 2000, see also Conway \et\ 1971; Roberts \et\ 1975). There is, however, no correlation between the degree of circular and linear polarization.

The circularly polarized emission from flat-spectrum sources is mainly from the compact core region.  The emission from their extended jets/ejecta is more linearly polarized (Homan \& Wardle 1999).  Observations of the radio flaring in 3C~273 between 1994 and 1997 show that the linear polarization increases only after the peak of the flare and when total flux density had decreased substantially (Stevens \et\ 1998).

In this sense, the polarization behaviour in GRO~J1655$-$40 and 3C~273 are similar.  In both cases the circular polarization originates from a compact region.  Moreover, the linear polarization increases only in the later stage of the flare.

Wu \et\ (2002) and Stevens \et\ (2002) suggest that the temporal and spectral evolution of GRO~J1655$-$40 and compact flat-spectrum radio sources can be explained in a unified framework based on the generalized-shock model of Marscher \& Gear (1985).  If the circularly polarized emission indeed originates from a compact region where the shocks are developed, the physical conditions in the environments near the compact object should play an important role
for the emission and radiative transport processes responsible for the circular polarization.
In particular, the intense radiation from the accretion disk could cool the relativistic electrons accelerated by the shocks, and hence determine the early stages of the shock evolution.
This in turn would determine the source spectral evolution, as well as the polarization properties at the onset of the outburst.  These issues will be discussed more quantitatively in Macquart \& Wu (in preparation).

\section{Conclusions}
We report the detection of circular polarization in the Galactic X-ray binary GRO~J1655$-$40.
The circular polarization changes sign at the lowest observed frequency, 1.4\,GHz during the commencement of the outburst.  This coincides with the evolution of the spectrum from being inverted near this frequency to a pure power law.  As such, it suggests that the circular polarization changes sign and peaks as the synchrotron opacity decreases to be comparable to unity.  
Such behaviour suggests that the circular polarization originates due to polarization conversion in the relativistic plasma associated with the source.

\begin{acknowledgements}
K.W. acknowledges the support from Australian Research Council through an ARC Australian Research Fellowship.   DCH acknowledges the support of a PPARC postdoctoral research grant to the University of Southampton and financial support from the Academy of Finland.  The Australia Telescope is funded by the Commonwealth Government of Australia for operation as a national facility by the CSIRO.  The authors thank Rob Fender for comments on the paper.
\end{acknowledgements}

\end{document}